# Phase-transition-induced giant Thomson effect for thermoelectric cooling

Rajkumar Modak[1,a),b)], Masayuki Murata[2,b)], Dazhi Hou[3], Asuka Miura[1,c)], Ryo Iguchi[1], Bin Xu[4], Rulei Guo[4], Junichiro Shiomi[4], Yuya Sakuraba[1,5], and Ken-ichi Uchida[1,4,6,7,a)]

[1]National Institute for Materials Science, Tsukuba 305-0047, Japan

[2]National Institute of Advanced Industrial Science and Technology, Tsukuba 305-8568, Japan

[3]ICQD, Hefei National Laboratory for Physical Sciences at Microscale, University of Science and Technology of China, Hefei, Anhui 230026, China

[4]Department of Mechanical Engineering, The University of Tokyo, Tokyo 113-8656, Japan

[5]PRESTO, Japan Science and Technology Agency, Saitama 332-0012, Japan

[6]Institute for Materials Research, Tohoku University, Sendai 980-8577, Japan

[7]Center for Spintronics Research Network, Tohoku University, Sendai 980-8577, Japan

a)**Authors to whom correspondence should be addressed:** MODAK.Rajkumar@nims.go.jp and UCHIDA.Kenichi@nims.go.jp

b)**These authors contributed equally:** Rajkumar Modak, Masayuki Murata

c)**Present address:** Integrated Research for Energy and Environment Advanced Technology, Kyushu Institute of Technology, Kitakyushu, Fukuoka 804-8550, Japan

**ABSTRACT**

The Seebeck and Peltier effects have been widely studied and used in various thermoelectric technologies, including thermal energy harvesting and solid-state heat pumps. However, basic and applied studies on the Thomson effect, another fundamental thermoelectric effect in conductors, are limited despite the fact that the Thomson effect allows electronic cooling through the application of a temperature gradient bias rather than the construction of junction structures. In this article, we report the observation of a giant Thomson effect that appears owing to magnetic phase transitions. The Thomson coefficient of FeRh-based alloys reaches large values approaching $-1{,}000$ $\mu$VK$^{-1}$ around room temperature because of the steep temperature dependence of the Seebeck coefficient associated with the antiferromagnetic-ferromagnetic phase transition. The Thomson coefficient is several orders of magnitude larger than the Seebeck coefficient of the alloys. Using the active thermography technique, we demonstrate that the Thomson cooling can be much larger than Joule heating in the same material even in a nearly steady state. The operation temperature of the giant Thomson effect in the FeRh-based alloys can be tuned over a wide range by applying an external magnetic field or by slightly changing the composition. Our findings provide a new direction in the materials science of thermoelectrics and pave the way for thermal management applications using the Thomson effect.

## I. INTRODUCTION

The Thomson effect refers to the release or absorption of heat as a result of applying a charge current to a conductor under a temperature gradient. The effect was predicted by William Thomson, who later became known as Lord Kelvin.[1,2] The Thomson effect is one of the nonlinear thermoelectric phenomena; in contrast to the linear-response thermoelectric phenomena, *e.g.*, the Seebeck and Peltier effects, the heat production rate induced by the Thomson effect is proportional to both the applied charge current density $\mathbf{j}_c$ and temperature gradient $\nabla T$:

$$\dot{q} = -\tau \mathbf{j}_c \cdot \nabla T \qquad (1)$$

where $\tau$ is the Thomson coefficient. Thus, when $\tau$ is finite, the Thomson effect enables thermoelectric cooling/heating in homogenous materials (Fig. 1), which is in stark contrast to conventional Peltier devices that require junction structures comprising different materials. Despite its technological potential and scientific importance,[3–5] the fundamental physics, materials science, and thermoelectric applications of the Thomson effect remain to be investigated. To address this gap, direct imaging measurements of the temperature modulation induced by the Thomson effect have recently been reported.[6] This versatile measurement technique makes it possible to investigate the detailed behaviors and functionalities of this phenomenon.

Materials with large Thomson coefficients are indispensable for realizing thermoelectric applications of the Thomson effect. In a conductor, $\tau$ obeys the first Thomson (or Kelvin) relation

$$\tau = T \frac{dS}{dT} \qquad (2)$$

where $S$ is the Seebeck coefficient and $T$ is the absolute temperature. Equation (2) indicates that materials with a sharp temperature dependence of $S$ are potential candidates to show large cooling/heating induced by the Thomson effect. However, the temperature derivative of $S$, that is, $\tau$, is usually very small. Although recent experiments have shown that $\tau$ in a $Bi_{88}Sb_{12}$ alloy is strongly enhanced under a magnetic field,[6] the obtained $\tau$ value is still below 100 μVK$^{-1}$. The small magnitude of $\tau$ hinders research on the Thomson effect.

In this work, we demonstrate that the Thomson coefficient can be much larger than the Seebeck coefficient with the aid of magnetic phase transitions. We focus on FeRh-based alloys, which show a sharp temperature dependence of $S$ associated with the antiferromagnetic-ferromagnetic phase transition.[7–11] By precisely controlling their composition, the phase transition temperature can be adjusted to room temperature. The determined $\tau$ for the $Fe_{49.0}Rh_{50.8}Ni_{0.2}$ alloy at room temperature reached −816 μVK$^{−1}$ (−906 μVK$^{−1}$) during the warming (cooling) cycle, which is nearly two orders of magnitude larger than $S$ of the same material. Using a recently developed thermoelectric imaging technique,[12–16] we verified the large Thomson-effect-induced temperature modulation in the $Fe_{49.0}Rh_{50.8}Ni_{0.2}$ alloy with a temperature gradient bias and demonstrated that the Thomson cooling can be much larger than Joule heating (Fig. 1). This demonstration reveals the potential of nonlinear thermoelectric effects and provides unconventional concepts for designing thermal management devices that do not require junction structures.

## II. RESULTS AND DISCUSSION
### A. Estimation of Thomson coefficient of Ni-doped FeRh alloy

FeRh-based alloys are well-known materials that exhibit a first-order magnetic phase transition from the antiferromagnetic (ferromagnetic) to ferromagnetic (antiferromagnetic) state with increasing (decreasing) $T$. Over the past decades, FeRh-based alloys have become popular among researchers owing to their unique physical properties governed by this phase transition. These alloys can be utilized in various applications, such as magnetic refrigeration,[17–20] heat-assisted magnetic recording,[21,22] and multiferroic devices.[23–27] Among the FeRh-based alloys, we chose compositions close to $Fe_{49}Rh_{51}$, which shows the magnetic phase transition just above room temperature.[17–20] To adjust the transition temperature to the desired room temperature, a very small amount of Ni (~0.2 %) was doped into the $Fe_{49}Rh_{51}$ alloy (see Methods for sample preparation). The phase transition of the $Fe_{49.0}Rh_{50.8}Ni_{0.2}$ alloy was confirmed by measuring the $T$ dependence of magnetization. As shown in Fig. 2(a), the alloy exhibits the antiferromagnetic-to-ferromagnetic (ferromagnetic-to-antiferromagnetic) transition around 310 K (300 K) during the warming (cooling) cycle; the difference in the phase transition temperature between the warming and cooling cycles is attributed to the hysteresis of the first-order phase transition.

Figure 2(c) shows the $T$ dependence of $S$ for the $Fe_{49.0}Rh_{50.8}Ni_{0.2}$ alloy during the warming and cooling cycles. The alloy has different $S$ values between the antiferromagnetic and ferromagnetic states and shows a sharp change around the temperatures across the phase transition [compare Fig. 2(c) with 2(a)]. As can be seen from Eq. (2), such a sharp $S$ change is desirable for obtaining large $\tau$. Figure 2(d) shows the $\tau$ values calculated from the $T$ dependence of $S$. The $\tau$ value of the $Fe_{49.0}Rh_{50.8}Ni_{0.2}$ alloy reached $-816$ µVK$^{-1}$ at $T = 312$ K ($-906$ µVK$^{-1}$ at $T = 305$ K) during the warming (cooling) cycle, which is 91 times (82 times) larger than the $S$ values of the alloy at the same temperature. The observed $\tau$ values are the largest reported to date and much larger than the previously reported values for the $Bi_{88}Sb_{12}$ alloy as well as other materials including a widely-used thermoelectric material,[6,28,29] although $S$ of the $Fe_{49.0}Rh_{50.8}Ni_{0.2}$ alloy is much smaller than that of these materials (Table I). This result indicates that the giant Thomson effect can appear in phase transition materials despite their small $S$. We confirmed that the electrical resistivity (thermal conductivity) of the $Fe_{49.0}Rh_{50.8}Ni_{0.2}$ alloy exhibits a small change (almost no change) across the phase transition temperature [Fig. 2(b)].

**B. Direct measurement of Thomson-effect-induced temperature modulation**

To directly observe the temperature modulation induced by the giant Thomson effect in the $Fe_{49.0}Rh_{50.8}Ni_{0.2}$ alloy, we used the thermoelectric imaging method based on lock-in thermography (LIT).[6,14–16] The experimental setup for the LIT measurement is shown in Fig. 3(a). A bar-shaped $Fe_{49.0}Rh_{50.8}Ni_{0.2}$ slab was bridged between two heat baths and a chip heater was attached to the center of the slab. A steady temperature gradient was generated along the $y$ direction by applying a charge current to the heater. In this configuration, the direction of $\nabla T$ is reversed between $y > 0$ (around region R1) and $y < 0$ (around region R2) [Fig. 3(a)], where $y = 0$ is set at the heater position. This $\nabla T$ distribution is useful for direct imaging measurements of the Thomson effect because the dependence of the heat release/absorption on the $\nabla T$ direction can be confirmed in a single measurement.[6] To perform the LIT measurements, a square-wave-modulated AC charge current with the amplitude $J_c$, frequency $f$, and zero DC offset was applied to the slab along the $y$ direction in the presence of a steady temperature gradient. The distribution of the periodic temperature change in response to the charge current at the top surface of the slab was recorded using an infrared camera. The obtained thermal images are transformed into lock-in amplitude $A$ and phase $\phi$ images through Fourier analysis, where $A$ represents the magnitude of the current-induced temperature change and $\phi$ the sign as well as the time delay of the temperature modulation. Under these conditions, the direction of the charge current is the same over the slab, whereas the direction of $\nabla T$ at R1 is opposite to that at R2 [Fig. 3(a)]. Therefore, the sign of the temperature modulation due to the Thomson effect is expected to be reversed between R1 and R2 [Fig. 3(b) and Eq. (1)]. During the LIT measurements, the sample was first heated by applying a large heater power of $P > 60$ mW and then the $P$ value was set in the order of 60, 45, 30, 15, and 0 mW. Thus, the thermal history of the sample resembles the cooling cycle in Fig. 2(d). The LIT measurements were performed at room temperature and atmospheric pressure.

In Fig. 3(c), we show the observed $A$ and $\phi$ images for the $Fe_{49.0}Rh_{50.8}Ni_{0.2}$ slab at $J_c = 300.0$ mA, $f = 1.0$ Hz, and $P = 60$ mW. The $A$ signals exhibit two peaks around R1 and R2 and that the $\phi$ difference between R1 and R2 is approximately 180°, indicating that the sign of the current-induced temperature modulation at R1 is opposite to that at R2. This behavior is consistent with Eq. (1), and confirms the presence of heating/cooling due to the Thomson effect. It should be noted that in this experimental configuration, the measured LIT images may include contributions not only from the Thomson effect but also from the Peltier effect generated at the ends of the sample connected to the electrodes.[6] In fact, we observed finite Peltier signals at $P = 0$ mW, that is, in the absence of a steady temperature gradient [Fig. 3(d)]. However, the magnitude of the Peltier-effect-induced temperature modulation in the present experiment is negligibly small as compared to the temperature modulation at $P = 60$ mW. We can thus regard the observed temperature modulation in Fig. 3(c) as the pure Thomson-effect-induced signals without any background subtraction. The spatial distribution of the temperature modulation in Fig. 3(c) appears as a consequence of the sign reversal of the Thomson signal across the center of the slab and the thermal connection of both ends of the slab to the heat baths.

To investigate the detailed behaviors of the Thomson effect in the $Fe_{49.0}Rh_{50.8}Ni_{0.2}$ alloy, we performed LIT measurements at different $P$ values. Figure 4(a) shows the steady-state temperature profiles along the $y$ direction on the top surface of the sample measured at $P = 60, 45, 30$, and $15$ mW. We confirmed that the temperature gradient bias monotonically increases with increasing $P$ and that the temperature distribution is not affected by the Joule heating due to the charge current applied to the sample. As shown in the $A$ profiles along the $y$ direction, the magnitude of the temperature modulation due to the Thomson effect increases with increasing $P$. We also found that the $A$ signals at R1 and R2 are proportional to $J_c$ [Fig. 4(b)]. These behaviors are in good agreement with the characteristics of the Thomson effect described in Eq. (1). In the inset to Fig. 4(b), we show the $f$ dependence of $A$ at R1 and R2 measured at $P = 60$ mW and $J_c = 300$ mA, where $A$ is normalized by the value at 5 Hz. This result shows that the $A$ signals gradually increase with decreasing $f$ and reach a nearly steady state around 0.1 Hz. In Fig. 4(c), we show $A$ normalized by the square-wave amplitude of the charge current density $j_c$ as a function of the temperature gradient along the $y$ direction $\nabla_y T$ at R1 and R2, where $\nabla_y T$ was estimated by fitting the steady-state temperature profiles at R1 and R2 with linear functions. The observed nonlinear $\nabla_y T$ dependence of $A/j_c$ can be explained by the fact that $\tau$ of the $Fe_{49.0}Rh_{50.8}Ni_{0.2}$ alloy exhibits the strong temperature dependence around room temperature [Fig. 2(d)] and that the base temperature of the sample changes slightly with the temperature gradient [Fig. 4(a) and inset to Fig. 4(c)]. The strong temperature dependence of $\tau$ also affects the distribution of the Thomson-effect-induced temperature modulation; as shown in Fig. 4(a), the magnitude of $A$ exhibits maximum values around R1 and R2 at $P = 60$ mW, while the peak positions gradually shift with decreasing $P$. Figure 4(d) confirms that the $T$ dependence of the Thomson-effect-induced temperature modulation normalized by the temperature gradient, $\eta = |A/j_c \nabla_y T|$, follows that of $\tau$ for the cooling cycle estimated from Fig. 2(d). These LIT measurements show that by optimizing the operation environment, the $Fe_{49.0}Rh_{50.8}Ni_{0.2}$ alloy exhibits a large temperature modulation owing to the phase-transition-induced Thomson effect.

## C. Demonstration of current-induced cooling

We are now in a position to compare the Thomson cooling with Joule heating. In a material with sufficiently large $\tau$, temperature modulation due to the Thomson effect compensates for Joule heating under the application of a charge current and temperature gradient and enables current-induced cooling without junction structures (Fig. 1), which cannot be realized with the Peltier effect. To directly demonstrate this functionality, we performed the LIT measurements while applying a nearly uniform temperature gradient to the $Fe_{49.0}Rh_{50.8}Ni_{0.2}$ slab along the $y$ direction, where a heater was attached to one end of the sample [Fig. 5(a)]. Here, an ON/OFF-modulated AC charge current with the ON-state current magnitude $-J_c$ ($J_c$) was applied to the sample in the $y$ direction for case 1 (case 2). When such an input current is applied, both the thermoelectric and Joule-heating signals appear in LIT images [Fig. 5(b)],[15,30] which is different from the situation in Fig. 3. Current-induced cooling is realized under this condition when the heat absorption due to the Thomson effect is larger than the heat release due to Joule heating. The LIT measurements were performed at $f = 0.1$ Hz because LIT images at low lock-in frequencies show nearly steady-state temperature distributions [Inset to Fig. 4(b)].

In Figs. 5(c)-5(e), we show the steady-state temperature $T$ images and the raw $A$ and $\phi$ images for the Fe$_{49.0}$Rh$_{50.8}$Ni$_{0.2}$ alloy measured by applying the ON-state current magnitude of $J_c$ = 100 mA at $f$ = 0.1 Hz. To estimate the Joule heating background without the Thomson effect contribution, we first recorded the LIT images in the absence of a temperature gradient bias (at $P$ = 0 W) with the same charge current condition as that in case 1. As shown in the left images in Figs. 5(c)-5(e), an almost uniform heating signal appeared due to Joule heating. In the LIT images measured by applying the temperature gradient bias (at $P$ = 1.5 W), we observed much larger temperature modulation signals [see the center and right images in Figs. 5(c)-5(e) and the line profiles in Figs. 5(f)-5(h)]. This drastic signal enhancement cannot be explained by the change in Joule heating because of the small change in the electrical resistivity of the Fe$_{49.0}$Rh$_{50.8}$Ni$_{0.2}$ alloy across the phase transition temperature [Fig. 2(b)]. Importantly, the sign of the temperature modulation under the temperature gradient bias was found to be reversed by reversing the direction of the charge current [see the $\phi$ difference of approximately 180° between cases 1 and 2 shown in Figs. 5(e) and 5(h)]. This result indicates that even in the raw LIT images without background subtraction, the Thomson effect provides the dominant contribution to the temperature modulation in the nearly steady-state condition. We note that the nonuniform distribution of $A$ and $\phi$ observed here can be qualitatively explained by the fact that $\tau$ of the Fe$_{49.0}$Rh$_{50.8}$Ni$_{0.2}$ alloy strongly depends on the sample temperature [Fig. 2(d)]; the spatially dependent Thomson-effect-induced temperature modulation reflects the local temperature in the presence of the temperature gradient. This demonstration showcases active heating/cooling control based on the Thomson effect in a single material without any junction structures.

### D. Operation temperature tuning

The operation temperatures of the phase-transition-induced giant Thomson effect can easily be tuned by applying an external magnetic field or by changing the composition. To verify this characteristic, we measured the $T$ dependence of $S$ in the same Fe$_{49.0}$Rh$_{50.8}$Ni$_{0.2}$ alloy during the warming cycle at different magnetic fields ranging from 0 T to 5 T [Fig. 6(a)]. Figure 6(b) shows the corresponding $\tau$ values calculated using Eq. (2). We found that the position of the $\tau$ peak monotonically decreases with increasing the magnetic field, although the magnitude of $\tau$ remains unchanged. The magnetic field dependence of the $\tau$ peak position is consistent with the field dependence of the first-order phase transition in similar alloys.[10,11] We also confirmed that the $\tau$ peak position shifts to higher (lower) temperatures with decreasing (increasing) Ni content in Ni-doped FeRh alloys. As shown in Figs. 6(c) and 6(d), the Fe$_{49.3}$Rh$_{50.7}$ alloy with no Ni content (the Fe$_{47.9}$Rh$_{51.1}$Ni$_{1.0}$ alloy with a higher Ni content) exhibits the maximum $\tau$ value of $-556$ μVK$^{-1}$ at $T$ = 342 K ($-726$ μVK$^{-1}$ at $T$ = 256 K) during the warming cycle. These features enable the flexible design of temperature modulators based on the phase-transition-induced Thomson effect.

### III. SUMMARY

In this work, we have reported the direct observation of the giant Thomson effect in the Ni-doped FeRh alloys originating from their antiferromagnetic-ferromagnetic phase transition. The alloys exhibit a large Thomson coefficient approaching $-1,000$ μVK$^{-1}$ around the phase transition temperatures, which is nearly two orders of magnitude larger than their Seebeck coefficient. Our results indicate that even substances with small $S$, which are not conventionally regarded as thermoelectric materials, can be used for thermoelectric conversion based on the Thomson effect if the temperature dependence of $S$ is large. This finding will invigorate materials science studies on nonlinear or higher-order thermoelectric phenomena and their applications for thermal energy engineering. Although we have demonstrated the potential of the Thomson effect as a thermoelectric converter using the FeRh-based alloys with specific compositions, the vast number of phase transition materials and the easy tunability of their phase transition temperatures offer highly flexible material choices for Thomson-effect-based thermoelectric devices at any desired operation temperatures. Our demonstrations provide a pathway for the construction of programmable thermal management devices in which the Thomson effect compensates Joule heating. Finally, we would also like to mention that, while practical applications of the Thomson device require the evaluation of its cooling efficiency, the coefficient of performance for the Thomson device has not been formulated so far. To do this, the effect of the two input energies to drive the Thomson device, i.e., a charge current and an external temperature gradient, needs to be taken into consideration in the coefficient of performance appropriately.[31] Thus, not only materials science studies but also fundamental physics studies are necessary to develop thermal management applications based on the Thomson effect.

## IV. METHODS

### A. Sample preparation

To prepare the undoped and Ni-doped FeRh alloy samples, pure Fe, Rh, and Ni shots with 99.99 % purity were first weighed in appropriate amounts and placed into an arc-melting chamber. Sufficiently large alloy ingots were then prepared by completely melting and mixing the metals in an Ar atmosphere. To ensure homogeneous mixing, each alloy was melted 10 times followed by annealing in a high-temperature vacuum furnace at 1000 °C for 72 h, and then slowly cooled to room temperature for 10 h. The alloy pieces used for various characterizations were cut from the same ingots using a diamond wire saw. Here, the post-annealing process is necessary to observe the giant Thomson effect in these alloys since the steep change in $S$ cannot be obtained in as-prepared alloys without the post-annealing due to the absence of clear first-order ferromagnetic-antiferromagnetic phase transition. The compositions of the alloys were evaluated using an inductively coupled plasma optical emission spectrometer and determined to be $Fe_{49.0}Rh_{50.8}Ni_{0.2}$, $Fe_{49.3}Rh_{50.7}$, and $Fe_{47.9}Rh_{51.1}Ni_{1.0}$.

### B. Measurement of magnetization

The $T$ dependence of the magnetization of the $Fe_{49.0}Rh_{50.8}Ni_{0.2}$ alloy in Fig. 2(a) was measured using a vibrating sample magnetometer. A slab with dimensions of $2.5 \times 2.5 \times 2.0$ mm$^3$ was used for the magnetization measurement. The data were recorded during the warming and cooling cycles under a constant magnetic field of 1 T along the 2.5-mm direction. We confirmed that the magnetization of the $Fe_{49.0}Rh_{50.8}Ni_{0.2}$ alloy in the ferromagnetic state was saturated at 1 T.

### C. Measurement of Seebeck coefficient and electrical resistivity

The $T$ dependence of $S$ was measured by the steady-state method. Here, we used a bar-shaped $Fe_{49.0}Rh_{50.8}Ni_{0.2}$ slab with a thickness of 0.48 mm, width of 1.54 mm, and length of 15.50 mm, a $Fe_{49.3}Rh_{50.7}$ slab with a thickness of 0.51 mm, width of 1.41 mm, and length of 15.03 mm, and a $Fe_{47.9}Rh_{51.1}Ni_{1.0}$ slab with a thickness of 0.46 mm, width of 1.50 mm, and length of 15.12 mm. Two Cu plates with approximate dimensions of $5 \times 5 \times 1$ mm$^3$ and a groove of the same width as the sample were fixed on both ends of the sample with Ag epoxy (Sk100SDN, Diemat). A small 120-Ω heater was attached to the hot-side Cu plate with a silicone adhesive (1225B, ThreeBond) to generate a temperature gradient. To measure the temperature difference, type-T differential thermocouples with a diameter of 25 μm were thermally attached to the Cu plates. The two Cu plates were fixed on the AlN substrate of a sample stage with Ag epoxy (Dotite AA55, Fujikura Kasei). The temperature of the sample stage was measured using a Cernox temperature sensor and controlled with a GM refrigerator and heater. The thermoelectric voltage measurements were conducted in a vacuum of less than $10^{-3}$ Pa. $S$ was determined by linier fitting of the measured thermoelectric voltage at the three small temperature differences of 0.2, 0.6 and 1.0 K. We confirmed that, at all the temperatures including the phase transition region, the measured thermoelectric voltage was linearly proportional to the applied temperature differences and the linear fitting of the thermoelectric voltage was not affected by small voltage offsets, where the coefficient of determination of the fitting is > 0.99 for all the data. This result validates the accurate estimation of $S$ even around the phase transition temperatures. During the measurement of $S$, the electrical resistivity of the samples was also measured by the four-probe method, where the distances between the voltage electrodes for the $Fe_{49.0}Rh_{50.8}Ni_{0.2}$, $Fe_{49.3}Rh_{50.7}$, and $Fe_{47.9}Rh_{51.1}Ni_{1.0}$ slabs were 3.19 mm, 2.93 mm, and 3.26 mm, respectively. The data in Fig. 6(a) were measured during the heating cycle by applying magnetic fields of 1 T, 3 T, and 5 T to the $Fe_{49.0}Rh_{50.8}Ni_{0.2}$ alloy along the width direction by using a superconducting magnet.

### D. Lock-in thermography

To measure the Thomson effect using LIT, we used a bar-shaped $Fe_{49.0}Rh_{50.8}Ni_{0.2}$ slab with lengths of 1.53 mm, 14.00 mm, and 0.50 mm along the $x$, $y$, and $z$ directions, respectively. The sample was bridged between two alumite-coated Al blocks, which functioned as heat baths, and fixed with a thermally conductive silicone adhesive (COM-G52, COM Institute). The ends of the sample were thermally connected, but electrically insulated from the Al blocks. For the experiments shown in Fig. 3 (Fig. 5), a chip heater with a resistance of 1 kΩ (100 Ω) and size of 3.2 × 0.4 × 1.6 mm$^3$ (4.2 × 0.7 × 3.0 mm$^3$) was attached to the center of the bottom surface (one end) of the sample; by applying a charge current to the heater, the temperature gradient of which the direction is reversed around the center (almost uniform temperature gradient) was generated. In the experiments shown in Fig. 3, the heater output was used only for generating the temperature gradient. In contrast, in the experiments shown in Fig. 5, the heater output was transferred not only to the sample but also to the Al block. Thus, the heater power applied during the LIT measurements in Fig. 5 ($P$ = 1.5 W) was much larger than that in Fig. 3 ($P \leq 60$ mW). If the Thomson-effect-induced temperature modulation is normalized by the temperature gradient, not the heat power, the magnitudes of the signals in both the experiments are comparable. We note that the temperature and magnetic field dependences of the heater resistance are negligibly small.[6] To enhance the infrared emissivity and ensure uniform emission properties, the top surface of the sample was coated with insulating black ink with an emissivity of > 0.95.

**ACKNOWLEDGMENTS**

The authors thank T. Chiba for valuable discussions and M. Isomura for technical supports. This work was supported by Grant-in-Aid for Scientific Research (B) (19H02585) from JSPS KAKENHI, Japan, CREST "Creation of Innovative Core Technologies for Nano-enabled Thermal Management" (JPMJCR17I1), from JST, Japan, and "Mitou challenge 2050" (P14004) from NEDO, Japan. A.M. was supported by JSPS through Research Fellowship for Young Scientists (18J02115).

**AUTHOR DECLARATIONS**
**Conflict of Interest**

The authors declare no competing interests.

**Author Contributions**

K.U. supervised the study, conceived the idea, and designed the experiments with input from D.H. R.M. prepared and characterized the samples. R.M. collected and analyzed the data except for the Seebeck coefficient with help from K.U., A.M., R.I., B.X., R.G., J.S, and Y.S. M.M. measured the temperature and magnetic field dependences of the Seebeck coefficient. R.M. and K.U. prepared the manuscript and developed an explanation of the experiments. All the authors discussed the results and commented on the manuscript.

## DATA AVAILABILITY

The data that support the findings of this study are available from the corresponding authors upon reasonable request.

## REFERENCES

[1] W. Thomson, Proc. R. Soc. Edinburgh **3**, 91 (1857).

[2] W. Thomson, Trans. R. Soc. Edinburgh **21**, 123 (1857).

[3] G. J. Snyder, E. S. Toberer, R. Khanna, and W. Seifert, Phys. Rev. B **86**, 045202 (2012).

[4] W. Seifert, G. J. Snyder, E. S. Toberer, C. Goupil, K. Zabrocki, and E. Müller, Phys. Status Solidi A **210**, 1407 (2013).

[5] Y. Amagai, T. Shimazaki, K. Okawa, H. Fujiki, T. Kawae, and N. H. Kaneko, AIP Adv. **9**, 065312 (2019).

[6] K. Uchida, M. Murata, A. Miura, and R. Iguchi, Phys. Rev. Lett. **125**, 106601 (2020).

[7] G. Shirane, C.W. Chen, P. A. Flinn, and R. Nathans, Phys. Rev. **131**, 183 (1963).

[8] M. Pugacheva, J. A. Morkowski, A. Jezierski, and A. Szajek, Solid State Commun. **92**, 731 (1994).

[9] J. Kudrnovský, V. Drchal, and I. Turek, Phys. Rev. B **91**, 014435 (2015).

[10] N. Pérez, A. Chirkova, K. P. Skokov, T. G. Woodcock, O. Gutfleisch, N. V. Baranov, K. Nielsch, and G. Schierning, Mater. Today Phys. **9**, 100129 (2019).

[11] N. Pérez, C. Wolf, A. Kunzmann, J. Freudenberger, M. Krautz, B. Weise, K. Nielsch, and G. Schierning, Entropy **22**, 244 (2020).

[12] O. Breitenstein, W. Warta, and M. Langenkamp, *Lock-in Thermography:Basics and Use for Evaluating Electronic Devices and Materials* (Springer, Berlin/Heidelberg, 2010).

[13] O. Wid, J. Bauer, A. Müller, O. Breitenstein, S. S. P. Parkin, and G. Schmidt, Sci. Rep. **6**, 28233 (2016).

[14] S. Daimon, R. Iguchi, T. Hioki, E. Saitoh, and K. Uchida, Nat. Commun. **7**, 13754 (2016).

[15] S. Daimon, K. Uchida, R. Iguchi, T. Hioki, and E. Saitoh, Phys. Rev. B **96**, 024424 (2017).

[16] K. Uchida, S. Daimon, R. Iguchi, and E. Saitoh, Nature **558**, 95 (2018).

[17] E. Stern-Taulats, A. Planes, P. Lloveras, M. Barrio, J. L. Tamarit, S. Pramanick, S. Majumdar, C. Frontera, and L. Mañosa, Phys. Rev. B **89**, 214105 (2014).

[18] A. Chirkova, K. P. Skokov, L. Schultz, N. V. Baranov, O. Gutfleisch, and T. G. Woodcock, Acta Mater. **106**, 15 (2016).

[19] E. Stern-Taulats, T. Castán, A. Planes, L. H. Lewis, R. Barua, S. Pramanick, S. Majumdar, and L. Mañosa, Phys. Rev. B **95**, 104424 (2017).

[20] A. Gràcia-Condal, E. Stern-Taulats, A. Planes, and L. Mañosa, Phys. Rev. Mater. **2**, 084413 (2018).

[21] J. U. Thiele, S. Maat, and E. E. Fullerton, Appl. Phys. Lett. **82**, 2859 (2003).

[22] P. -W. Huang and R. H. Victora, IEEE Trans. Magn. **50**, 8600204 (2014).

[23] P. A. Algarabel, M. R. Ibarra, C. Marquina, A. Del Moral, J. Galibert, M. Iqbal, and S. Askenazy, Appl. Phys. Lett. **66**, 3061 (1995).

[24] R. O. Cherifi, V. Ivanovskaya, L. C. Phillips, A. Zobelli, I. C. Infante, E. Jacquet, V. Garcia, S. Fusil, P. R. Briddon,

N. Guiblin, A. Mougin, A. A. Ünal, F. Kronast, S. Valencia, B. Dkhil, A. Barthélémy, and M. Bibes, Nat. Mater. **13**, 345 (2014).

[25] I. Fina, A. Quintana, X. Martí, F. Sánchez, M. Foerster, L. Aballe, J. Sort, and J. Fontcuberta, Appl. Phys. Lett. **113**, 152901 (2018).

[26] P. Dróżdż, M. Ślęzak, K. Matlak, B. Matlak, K. Freindl, D. Wilgocka-Ślęzak, N. Spiridis, J. Korecki, and T. Ślęzak, Phys. Rev. Appl. **9**, 034030 (2018).

[27] K. Tanaka, T. Moriyama, T. Usami, T. Taniyama, and T. Ono, Appl. Phys. Express **11**, 013008 (2018).

[28] N. F. Hinsche, F. Rittweger, M. Hölzer, P. Zahn, A. Ernst, and I. Mertig, Phys. Status Solidi A **213**, 672 (2016).

[29] I. T. Witting, T. C. Chasapis, F. Ricci, M. Peters, N. A. Heinz, G. Hautier, and G. J. Snyder, Adv. Electron. Mater. **5**, 1800904 (2019).

[30] K. Uchida, M. Sasaki, Y. Sakuraba, R. Iguchi, S. Daimon, E. Saitoh, and M. Goto, Sci. Rep. **8**, 16067 (2018).

[31] T. Chiba *et al.*, "Coefficient of performance for thermoelectric Thomson device" (unpublished).

**TABLE I.** The Thomson coefficient $\tau$ and Seebeck coefficient $S$ of various materials near room temperature.

| Material | $S$ ($\mu$VK$^{-1}$) | $\tau$ ($\mu$VK$^{-1}$) | Ref. |
|---|---|---|---|
| Fe$_{49.0}$Rh$_{50.8}$Ni$_{0.2}$ | −11 | −906 | This work |
| Ni | −22 | −16 | 6 |
| Bi$_{88}$Sb$_{12}$ at $\mu_0 H$ = 0 T | −91 | 45 | 6 |
| Bi$_{88}$Sb$_{12}$ at $\mu_0 H$ = 0.9 T | −110 | 98 | 6 |
| $p$-type Bi$_2$Te$_3$ | 110-220 | ~150 | 28, 29 |

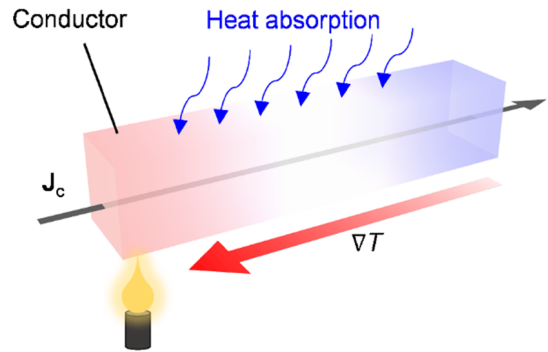

**FIG. 1.** Schematic of thermoelectric cooling based on Thomson effect. When a charge current **J**$_c$ and a temperature gradient $\nabla T$ are applied to a conductor, heat is released or absorbed depending on the relative direction of **J**$_c$ and $\nabla T$ as well as the sign of the Thomson coefficient $\tau$; the heat release/absorption due to the Thomson effect is proportional to $\tau$ and **J**$_c \cdot \nabla T$. Current-induced cooling is realized when the Thomson-effect-induced heat absorption is larger than the Joule-heating-induced heat release.

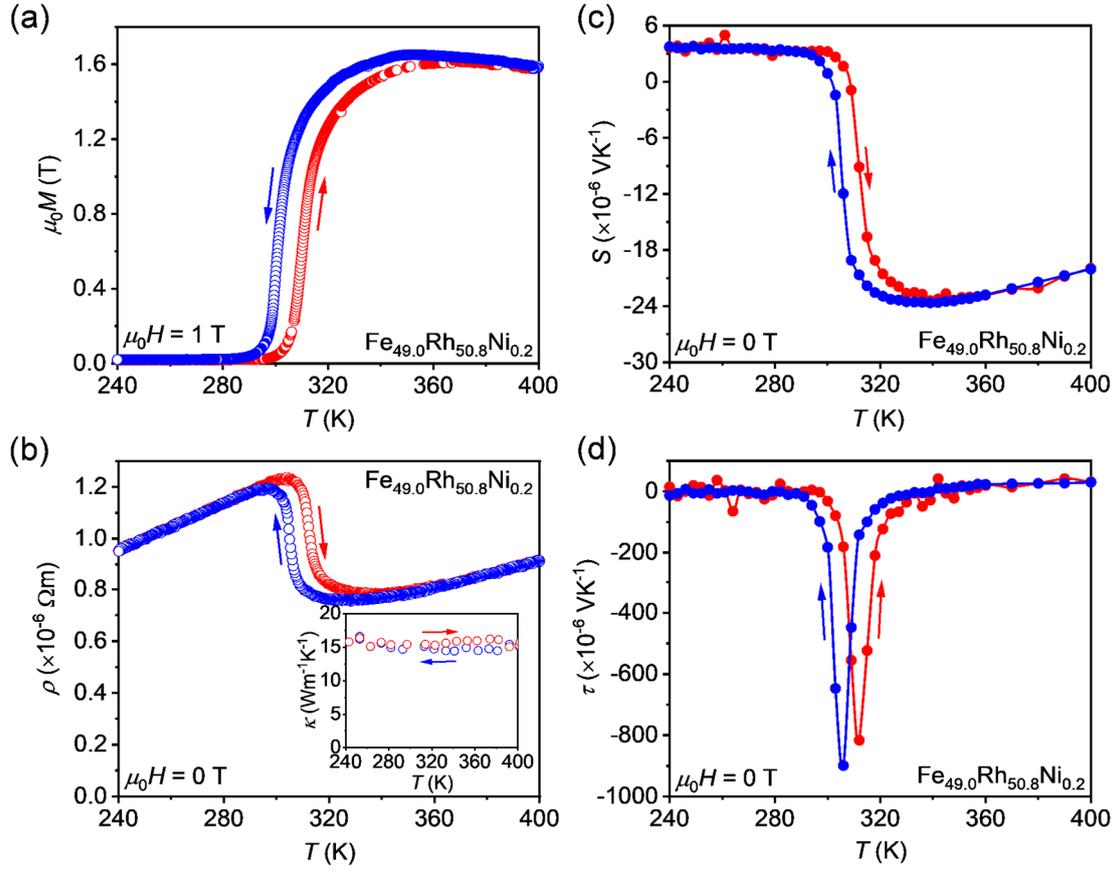

**FIG. 2.** Magnetic and transport properties. (a) Temperature $T$ dependence of the magnetization $M$ of the $Fe_{49.0}Rh_{50.8}Ni_{0.2}$ slab at a static external magnetic field of $\mu_0 H = 1$ T, measured during the warming (red) and cooling (blue) cycles. $\mu_0$ is the vacuum permeability. (b) $T$ dependence of the electrical resistivity $\rho$ of the $Fe_{49.0}Rh_{50.8}Ni_{0.2}$ slab at $\mu_0 H = 0$ T, measured during the warming (red) and cooling (blue) cycles. The inset to (b) shows the $T$ dependence of the thermal conductivity $\kappa$. The $\kappa$ values were estimated from the thermal diffusivity measured using the laser flash method, specific heat capacity measured using the differential scanning calorimetry, and density measured using the Archimedes method, where the contribution of latent heat to the specific heat capacity was subtracted. (c) $T$ dependence of the Seebeck coefficient $S$ of the $Fe_{49.0}Rh_{50.8}Ni_{0.2}$ slab at $\mu_0 H = 0$ T, measured during the warming (red) and cooling (blue) cycles. (d) $T$ dependence of the Thomson coefficient $\tau$ at $\mu_0 H = 0$ T estimated from the data in (c) based on Eq. (2). The solid curves connecting the data points in (c) and (d) were obtained using the Akima spline interpolation.

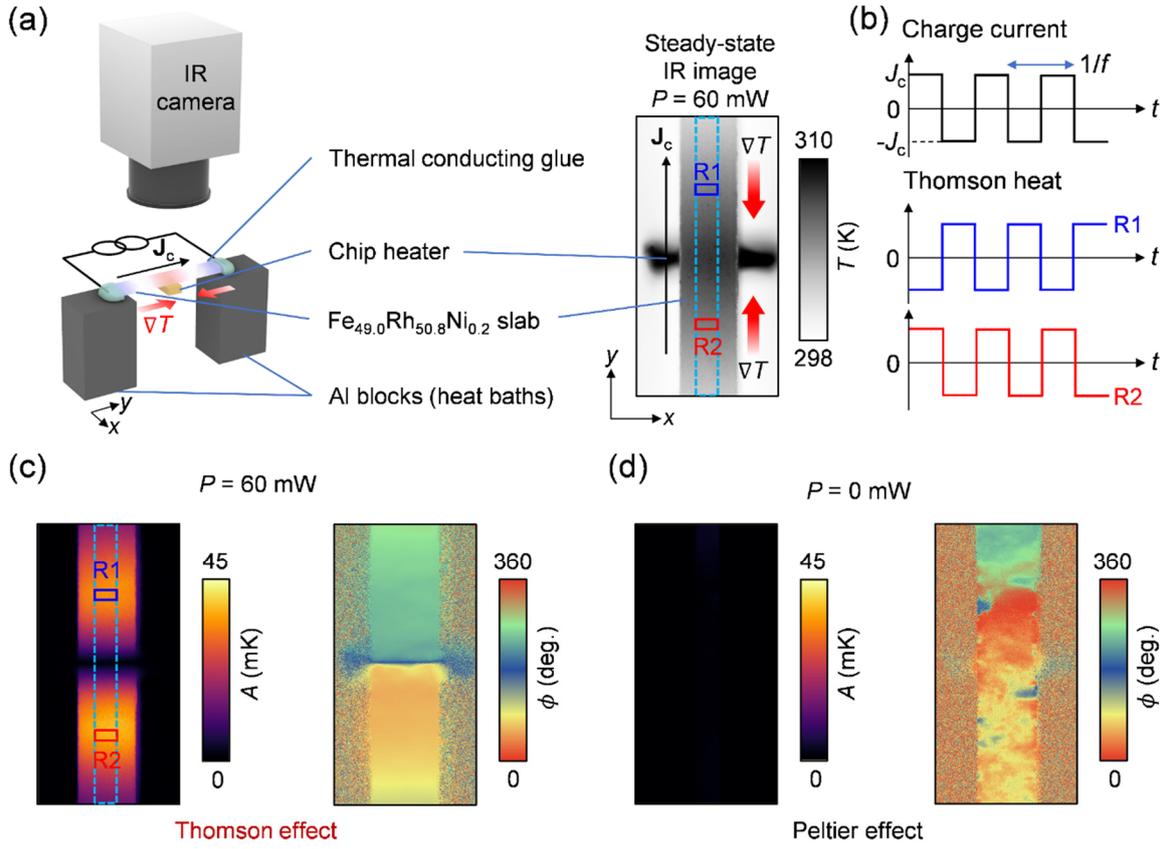

**FIG. 3.** Thermal imaging of Thomson-effect-induced temperature modulation. (a) Schematic of the experimental configuration for the measurements of the Thomson effect in the $Fe_{49.0}Rh_{50.8}Ni_{0.2}$ slab using the lock-in thermography (LIT) method and the steady-state $T$ image with applied $\nabla T$. $P$ denotes the heater power. (b) Time $t$ charts of the input charge current and output Thomson heat at the regions R1 and R2, defined by the blue and red rectangles in the $T$ image in (a), respectively. Here, a square-wave-modulated AC charge current with the amplitude $J_c$, frequency $f$, and zero DC offset was applied to the sample. (c), (d) Lock-in amplitude $A$ and phase $\phi$ images for the $Fe_{49.0}Rh_{50.8}Ni_{0.2}$ slab at $J_c$ = 300 mA, $f$ = 1.0 Hz, and $P$ = 60 mW (c) and 0 mW (d).

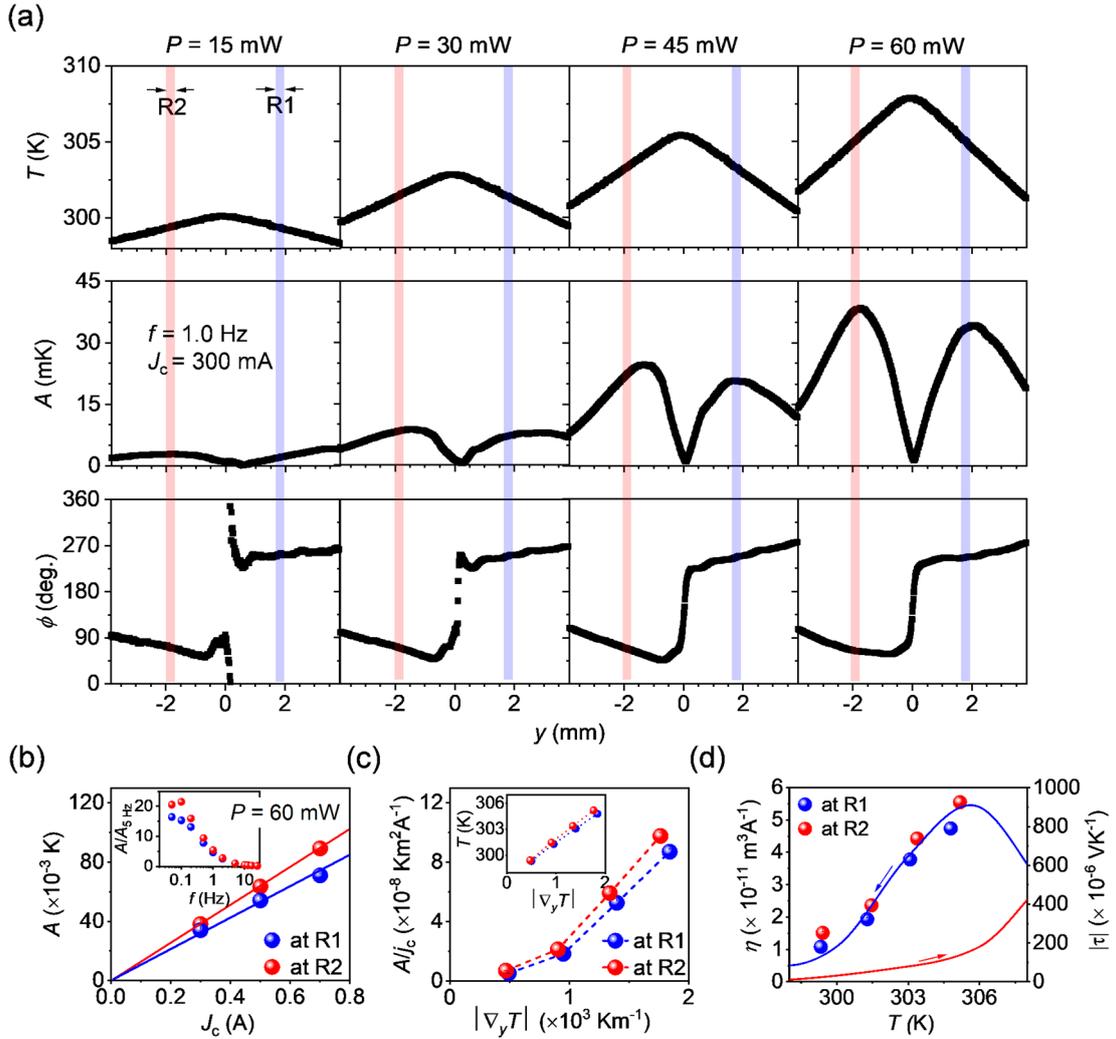

**FIG. 4.** Systematic measurements of Thomson-effect-induced temperature modulation. (a) $T$, $A$, and $\phi$ profiles along the $y$ direction at $f = 1.0$ Hz for various values of $P$. The areas used for calculating the profiles are marked by the blue dotted rectangles in Figs. 3(a) and 3(c), where the profiles were obtained by averaging 40 $y$-directional profiles along the $x$ direction. (b) $J_c$ dependence of $A$ at R1 and R2 at $f = 1.0$ Hz and $P = 60$ mW. The data points in (b) were obtained by averaging the values on the areas defined by the rectangles in Fig. 3(c), which have a size of 40 × 20 pixels (0.60 × 0.30 mm$^2$). The error bars represent the standard deviation of the data in the corresponding rectangles, which are smaller than the size of the data points. The inset to (b) shows the $f$ dependence of $A$ normalized by the value at 5 Hz $A_{5\,Hz}$ at R1 and R2, measured at $P = 60$ mW and $J_c = 300$ mA. (c) Dependence of $A/j_c$ at R1 and R2 on the temperature gradient along the $y$ direction $|\nabla_y T|$ at $f = 1.0$ Hz. $j_c$ denotes the charge current density. The inset to (c) shows the $|\nabla_y T|$ dependence of $T$ at R1 and R2; by increasing the temperature gradient, the base temperature also increases. (d) $T$ dependence of the Thomson signal $\eta$ ($= |A/j_c \nabla_y T|$) at $f = 1.0$ Hz and the absolute value of the Thomson coefficient $|\tau|$.

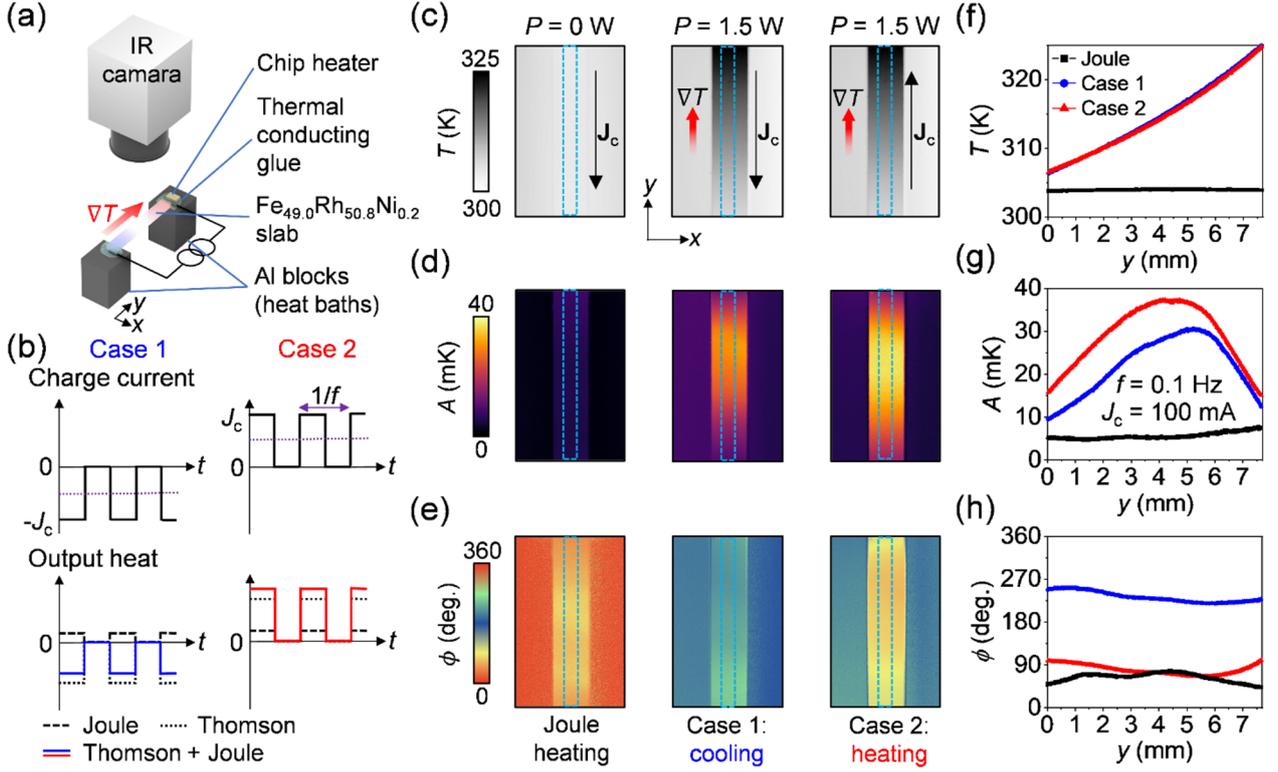

**FIG. 5.** Demonstration of Thomson cooling. (a) Schematic of the experimental configuration for demonstrating current-induced cooling due to the Thomson effect in the $Fe_{49.0}Rh_{50.8}Ni_{0.2}$ slab. (b) $t$ charts of the input charge current and output heat for cases 1 and 2. In case 1 (2), the ON/OFF-modulated AC charge current with the ON-state current magnitude $-J_c$ ($J_c$) and frequency $f$ was applied to the sample. The total output heat is determined by the summation of the Thomson effect and Joule heating contributions. The sign of the Thomson-effect-induced heat is reversed when the DC offset of the charge current is reversed, while Joule heating is independent of the current direction. (c)-(e) $T$, $A$, and $\phi$ images for the $Fe_{49.0}Rh_{50.8}Ni_{0.2}$ slab at $J_c = 100$ mA and $f = 0.1$ Hz in the conditions for measuring the Joule heating contribution (at $P = 0$ W) and the total temperature modulation for cases 1 and 2 (at $P = 1.5$ W). (f)-(h) $T$, $A$, and $\phi$ profiles along the $y$ direction in the conditions for measuring the Joule heating contribution (at $P = 0$ W; black lines) and total temperature modulation for the cases 1 and 2 (at $P = 1.5$ W; blue and red lines, respectively). The areas used for calculating the profiles are marked by the blue dotted rectangles in (c)-(e), where the profiles were obtained by averaging 40 $y$-directional profiles along the $x$ direction.

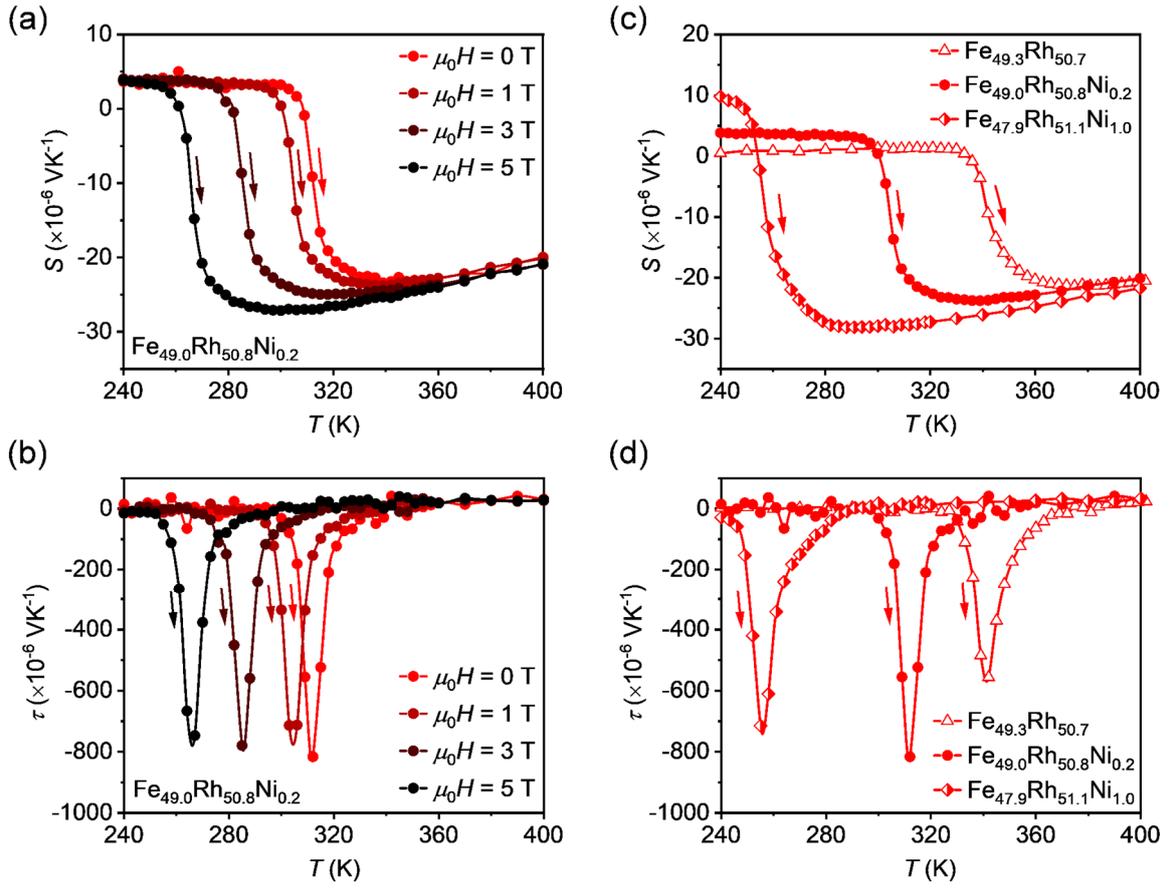

**FIG. 6.** Magnetic field and composition dependences. (a) $T$ dependence of $S$ of the $Fe_{49.0}Rh_{50.8}Ni_{0.2}$ slab at various values of $\mu_0 H$, measured during the warming cycle. (b) $T$ dependence of $\tau$, estimated from the data in (a) based on Eq. (2). (c) $T$ dependence of $S$ for the $Fe_{49.0}Rh_{50.8}Ni_{0.2}$ (circles), $Fe_{49.3}Rh_{50.7}$ (triangles), and $Fe_{47.9}Rh_{51.1}Ni_{1.0}$ (diamonds) slabs at $\mu_0 H = 0$ T, measured during the warming cycle. (d) $T$ dependence of $\tau$ estimated from the data in (c). The solid curves connecting the data points in (a)-(d) were obtained using the Akima spline interpolation.